# Comment on "Quantifying Long-term Scientific Impact"

*Jin-Li Guo, Qi Suo*

*Business School, University of Shanghai for Science and Technology, Shanghai 200093, China*

*e-mail:phd5816@163.com*

**Abstract:** The paper comments on "Quantifying long-term scientific impact". It indicates that there is a mistake of [D. S. Wang, C. Song, A. L. Barabási. Quantifying long-term scientific impact, Science, 342, 127 (2013)].

The number of papers published each year increases exponentially, in analogy to Moore's law describing the development of technology, indicating

$$N(t) \sim \exp(\beta t),  \quad (S1)$$

where $\beta$ is a constant in supplementary materials [1].

Barabási *et al.* propose a minimal citation model to quantify long-term scientific impact. In their model, the citation probability $\Pi_i$ of each old paper $i$ is supposed to satisfy Eq. (S2) in supplementary materials [1], that is,

$$\Pi_i(\Delta t) \sim \eta_i c_i^t P(\Delta t_i). \quad (S2)$$

As $\beta$ and $A$ are system wide parameters in supplementary materials [1], Barabási *et al.* use $\lambda_i = \eta_i \beta / A$ the relative fitness for each paper $i$. They obtained a conclusion that the ultimate citation of paper $i$ is

$$c_i^\infty = m(e^{\lambda_i} - 1). \quad (S14)$$

However, there is an error in the process of solving the model. In supplementary materials [1], when plugging $c_i^t = m(f(\eta_i, \Delta t_i) - 1)$ into Eq. (S5), Barabási *et al.* obtain an incorrect Eq. (S6). The Eq. (S6) in supplementary materials [1] should be

$$\frac{df(\eta_i, \Delta t_i)}{d\Delta t_i} = \frac{\beta N(t) \eta_i (f(\eta_i, \Delta t_i) - 1) P(\Delta t_i)}{\sum_{i=1}^{N} \eta_i (f(\eta_i, \Delta t_i) - 1) P(\Delta t_i)}, \quad (C1)$$

The solution of this equation, with the initial condition $f(\eta_i, 0) = 1$, is

$$f(\eta_i, \Delta t_i) = 1. \quad (C2)$$

Thus

$$c_i^t = m(f(\eta_i, \Delta t_i) - 1) = 0, \quad (C3)$$

From Eq. (C3), we get

$$c_i^\infty = 0. \quad (C4)$$

The formula predicts the ultimate citation ($c_i^\infty$) of paper $i$ is zero. That is to say, any paper will never be cited. Eq. (C4) also contradicts with Eq. (S14) in supplementary materials [1]. The result shows that there is a mistake of Ref.[1].

**Acknowledgement**

This paper is supported by the National Natural Science Foundation of China (Grant No.70871082) and the Shanghai First-class Academic Discipline Project (Grant No. S1201YLXK).